\documentclass[conference]{IEEEtran}
\IEEEoverridecommandlockouts
\usepackage{cite}
\usepackage{amsmath,amssymb,amsfonts}
\usepackage{algorithm}
\usepackage{algpseudocode}
\usepackage{graphicx}
\usepackage{textcomp}
\usepackage{xcolor}
\usepackage{subfigure}
\usepackage{multirow}
\usepackage{url}
\usepackage{booktabs}
\def\BibTeX{{\rm B\kern-.05em{\sc i\kern-.025em b}\kern-.08em
    T\kern-.1667em\lower.7ex\hbox{E}\kern-.125emX}}
    
\begin{document}

\title{Vehicle-to-Grid Technology meets Packetized Energy Management: A Co-Simulation Study}

%

\author{
\IEEEauthorblockN{Freddy Tuxworth and Adnan Aijaz}
\IEEEauthorblockA{
\text{Bristol Research and Innovation Laboratory, Toshiba Europe Ltd., United Kingdom }\\
firstname.lastname@toshiba-bril.com}\\
\vspace{-4em}
} 

%


\maketitle

\begin{abstract}
The global energy landscape is experiencing a significant transformation driven by increased awareness of climate change and rapid technological advancements in  renewable energy and electric vehicles (EVs). Packetized energy management (PEM) schemes are gaining attention as a potential solution for power management for effective load control. This study presents the development of a co-simulation platform to investigate  integration of vehicle-to-grid (V2G) with packetized energy trading (PET) in microgrid scenarios. The platform facilitates the interaction between EVs and prosumers, with a focus on responsive loads, and solar photovoltaic (PV) as intermittently available resources. Using the developed co-simulation, this study evaluates how V2G-capable EVs can enhance the stability and efficiency of PET-based microgrids. The results demonstrate the capability of V2G EVs to act as an energy reservoir, effectively managing demand-side load, thus mitigating its fluctuation from available supply while maintaining quality-of-service. 
\end{abstract}


\section{Introduction}
As climate change demands urgent action, integrating renewable energy into the power grid has become crucial for sustainability and reducing fossil fuel dependency. This transition enhances grid resilience but introduces challenges like fluctuating generation capacities, complicating energy management \cite{Safdar2013AMicro-grids}. Electrification of various sectors for decarbonization increases energy demands, requiring sophisticated grid and microgrid management strategies \cite{net_zero}.

\begin{figure}
    \centering
    \includegraphics[scale=0.34]{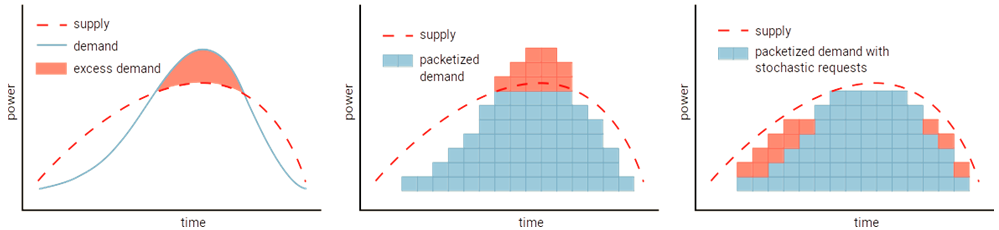}
    \caption{Illustration of the PEM concept (adapted from \cite{PEM_intro1}. The first graph shows a peak in demand overloading grid supply, the second shows the same load \emph{packetized}, and the third shows the same with excess demand shifted to times of higher supply availability.}
    \label{pem}
    \vspace{-1.5em}
\end{figure}

Packetized energy management (PEM) \cite{PEM_intro1} is a promising approach for demand-side load control. PEM represents energy in discrete time and load-bounded packets  which are requested from a controller. 
This facilitates granular control over energy distribution and utilization. Fig. \ref{pem} provides a simplified illustration of this concept.   The PEM concept has evolved over time and found particular utility in managing variable demand-side
loads such as heating, ventilation, and air conditioning (HVAC). 
PEM is considered under the umbrella of smart grid technologies and widely recognized as an enabler of the Energy Internet \cite{energy_internet}. 
The transition towards renewable energy sources and their integration into the power grid can be facilitated by PEM. 

Packetized energy trading (PET) applies packetization to energy markets, allowing energy trading from diverse sources via centralized or decentralized markets. 

On the other hand, vehicle-to-grid (V2G) technology enables electric vehicles (EVs) to contribute power back to the grid, providing various services including load balancing, power regulation, and reactive power support  \cite{V2G_benefits}.



\subsection{Related Work}
PEM has been the focus of various recent studies. 
Almassalkhi \emph{et al.}\cite{almassalkhi_duffaut_hines_frolik_paudyal_amini_2018} show potential utility of PEM in ensuring electric grid reliability using flexible and controllable distributed energy resources (DERs). Espinosa \emph{et al.}\cite{espinosa_khurram_almassalkhi_2021} show that PEM provides load balancing and ramping services for the grid through large-scale coordination of DERs. Brahma \emph{et al.}\cite{brahma_khurram_ossareh_almassalkhi_2022} propose a virtual battery model of PEM and a model predictive control framework. Zhang and Baillieul \cite{zhang_baillieul_2016} describe a new PEM framework, packetized direct load control, and emphasize trade-offs between controllability and consumer choice.  Almassalkhi \emph{et al.} \cite{PEM_intro1} explain how real-time coordination of demand through PEM can be an alternative to scaling up grid infrastructure. PET schemes have been investigated in some studies as well  \cite{RW_PET1}, \cite{RW_PET2}, particularly for energy trading in microgrids. 
Some studies have investigated  potential integration of V2G and smart grid technologies. Optimal EV charge scheduling has been studied in \cite{he_venkatesh_guan_2012, wu_aliprantis_ying_2012, rotering_ilic_2011}.  
The energy storage and computational capabilities of EVs are also promising for energy trading \cite{RW_EV_trading}.

\subsection{Objectives and Contributions}
This work centers on the idea that V2G-capable EVs, acting as mobile energy resources, can have a profound impact on the overall stability and efficiency of packetized energy microgrids. 
While previous studies have mostly explored V2G and PEM technologies separately, this work investigates the interaction between the two through co-simulation.
The objective is to explore the potential of PET schemes as a medium for the integration of EVs and renewable energy sources in microgrids, and to assess the impact on grid stability and energy optimization. The key contributions are as follows.

\begin{itemize}
\item We develop a co-simulation platform for PET providing seamless integration of EVs, renewable energy resources, household loads, and grid supply. Our platform significantly extends PEMT-CoSim \cite{PEMT}, a recently developed platform for PET, through a novel EV module and new market mechanism, among other enhancements. 

\item We establish quantitative metrics to objectively evaluate co-simulation framework in terms of energy optimization, grid stability, and integration efficiency.

\item We design and execute different co-simulation scenarios capturing interaction between different entities and power trading dynamics, and analyze the PET performance. 
\end{itemize}

\section{Overview of PEMT-CoSim}\label{sect_pemt}
PEMT-CoSim \cite{PEMT} provides an extensible co-simulation environment for integrating a variety of distinct, separately-simulated agents in a packetized energy system. It is built upon the transactive energy simulation platform (TESP) \cite{TESP}. PEMT-CoSim coordinates a collection of sub-simulation \emph{federates}, each of which
simulates a different aspect of the overall scenario. 
These federates interact and maintain simulation-time synchronization
using the HELICS broker \cite{helics}. The PEMT-CoSim federates are briefly described as follows. 

\textbf{HELICS Broker} -- HELICS is a co-simulation framework that enables synchronized communication between separate
domain-specific simulation federates. It is  employed for examining interactions between models and understanding the behavior of complex systems. The core of HELICS is a broker, which orchestrates exchange of value messages between federates and manages time synchronization.


\textbf{GridLAB-D} -- The GridLAB-D federate is responsible for modeling the houses, the loads they produce, and
the power flow of the microgrid. It is an integration of GridLAB-D, an existing open-source power distribution simulation software \cite{gridlabD}. Besides power flow solving and measurement, the key GridLAB-D feature used by PEMT-CoSim is the HVAC simulation, which models the change in air temperature over time of each house, considering external air temperature (provided by the weather federate) and the state of the HVAC system in that house. 

\begin{figure}[h]
    \centering
    \includegraphics[scale=0.24]{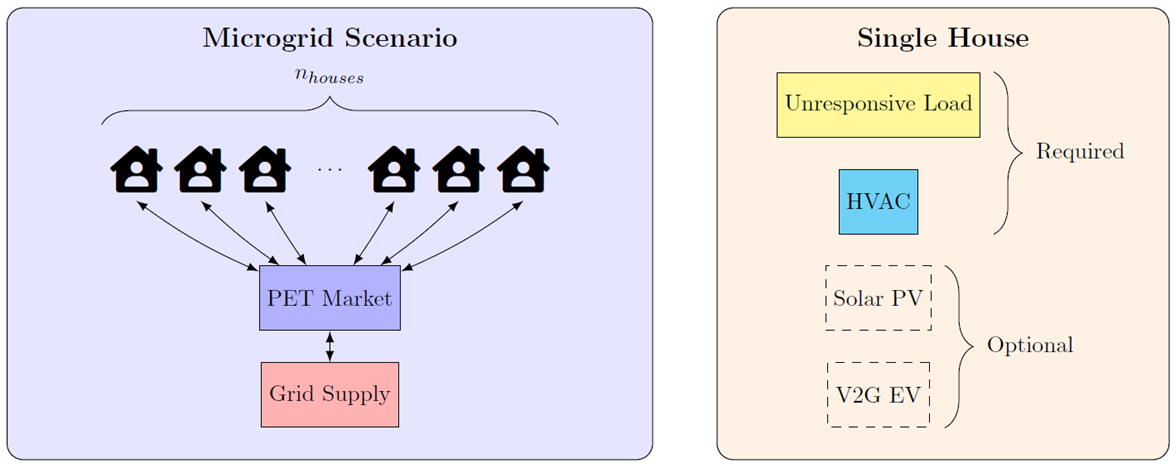}
    \caption{Illustration of overall system model and simulation scenario for PET.}
    \label{sim}
    \vspace{-1em}
\end{figure}

\textbf{Substation} -- The substation federate holds the business logic for  PET simulation. It has two key components: prosumer control and market simulation.
Each market round, it assigns each prosumer (house) a market role - buyer, seller, or non-participant - based on predicted load and solar photovoltaic (PV) supply (see Fig. \ref{sim}). Bids to buy or sell power in multiples of a defined packet quantity are formulated according to certain rules. The substation federate also implements a post-market control routine for each house. 
PEMT-CoSim employs a double auction market mechanism. Each bid submitted by a buyer/seller specifies the quantity of energy involved (in packets), the price at which the participant is willing to buy/sell, and whether the bid is unresponsive (part of the \emph{base load}) or responsive. The market then determines a single clearing price.
For bids that are marked as unresponsive, i.e., these cannot be rejected and form part of the base load, the market assigns them the maximum price, ensuring that they are always accepted.

\textbf{PyPower} -- The pertinent feature of the PyPower federate is that it provides the \emph{locational marginal price} (LMP)
of the grid supply to other federates. The LMP represents the marginal cost of supplying
the next increment of electric load at a specific location (in this case at the connection point between the microgrid and the wider grid distribution network). The PyPower federate is built around the PyPower Python library (www.github.com/rwl/PYPOWER). 


\textbf{Weather} -- The weather federate provides temperature, humidity, solar irradiance and wind speed data, which are
loaded from a dataset provided as part of TESP \cite{TESP}. 

\section{Proposed Co-simulation Framework}
\subsection{Experimental Design}
The overall system model and simulation scenario  is shown in Fig. \ref{sim}. It consists of community of \(n_{\textrm{houses}}\) houses, connected together in a microgrid. Every house is able to buy/sell
power in fixed-length packets by submitting bids to the power market. The microgrid is  connected to the wider grid power supply. 
The key characteristic of the simulated scenario is that all power consumed or supplied by
any agent connected to the microgrid must be accounted for by a transaction or transactions on the packetized market. This criterion enables the market to act as a power load regulator, since agents who wish to place a load onto the microgrid must first place a buy order and have it fulfilled. 
To assess the utility of the PET scheme and the contribution that EVs can make toward flattening the
grid load and minimizing excess demand, we compare several variations of the described scenario (see Section \ref{sec_cosim_res}).
Each simulation will be conducted for eight days of simulation time; the first four days of results will be discarded for accurate indication of the steady-state behaviour of the system.

\subsubsection{Houses, EVs, and Solar PV Arrays}

Each participating house in the PET system is equipped with standard electrical appliances and a cooling-only HVAC system. Among the $n_{houses}$, $n_{EV}$ houses have V2G-capable EVs and $n_{PV}$ have solar PV arrays. Household devices are categorized into \emph{responsive} (flexible) and \emph{unresponsive} (inflexible) loads. Unresponsive loads, such as appliances run on fixed schedules by the GridLAB-D federate, require constant power and their failure to receive it is a simulation failure. Responsive loads, like HVAC systems, and supplies, such as solar PV arrays, can adjust their operation based on signals from controllers. EVs act as both, switching between supplying or consuming power based on PET controller signals.

%

The EVs, which are capable of bidirectional charging/discharging only at home, and operate within realistic efficiency limits; charging at a rate slightly lower than their microgrid load and discharging slightly above their supply capability. The EV mobility patterns (explained later) dictate their availability for the microgrid, and act as traders in the PET scheme; they can only supply power when parked at home.


Solar PV arrays vary by house, with 8 to 20 panels each, capable of generating up to 480W per panel, based on data from TESP for PEMT-CoSim \cite{TESP}. These arrays adaptively contribute power depending on their solar capture, enhancing the microgrid's flexible energy resources.

\subsubsection{Grid Supply}
The grid power supply is a constant-capacity dispatchable power generation resource, which submits a fixed-quantity sell bid to the market every market round. The power supplied by the grid to the microgrid is adjusted based on the proportion of this bid which is fulfilled.
            The grid supply prices its sell orders at the LMP determined by a grid power flow simulation. The LMP is the cost of producing and supplying an additional unit of electricity at a particular node or location on the power grid. The LMP increases with the demand to supply ratio, although the specific mechanics that govern LMP determination are beyond the scope of this work.

\subsubsection{Market}
 The PET market operates as a price-first continuous double auction, where sell orders are treated as divisible (i.e. they may be partially fulfilled, requiring the seller to produce some fraction of the power they advertised for sale) and buy orders as indivisible (they must be either accepted or rejected in full). Note that this mechanic also allows for multiple sellers to fulfill a single buy order provided that their total sale quantity is at least the buy order quantity.
            Each market period $t_\text{market}$, the auction system aggregates bids from all traders, and then matches successively matches order orders with the aim of minimising the amount spent while fulfilling all buy orders possible.
           A $t_\text{market}$ of 300s, was adopted from PEMT-CoSim \cite{PEMT},  being short enough for traders to reasonably commit to, and long enough that simulations can complete in acceptable time.

\subsection{Bid Formulation}
At the start of each market period, houses must submit bids for desired power quantities for the upcoming period, involving \textit{power prediction} and \textit{price determination} processes.
Accurate power prediction ensures houses match actual needs, preventing shortfalls from underestimation or excess costs and grid issues from overestimation.
Pricing strategies strive to minimize energy costs while ensuring sufficient power. These interact with fluctuating power supplies and dynamic microgrid energy availability patterns, and necessitate fine-tuning to balance market demands and prevent grid overloads.

\subsubsection{Grid Supply}
The grid supply offers a fixed quantity of power in each market period, which is determined by a predetermined grid cap associated with each scenario. The bidding strategy for the grid supply's bids is straightforward; the price is always set at the LMP provided by the PyPower federate, and the bid quantity is always equal to the fixed grid load cap $P_\text{capacity}^\text{grid}$.

\subsubsection{House Loads}
The HVAC load is predicted using a simple rule - it is set to 4kW if the HVAC needs power due to deviation from the setpoint, or 0 if not. This fixed approximation of the average cooling load was found to be sufficiently precise. HVAC buy bids use a fixed price which is high, but lower than unresponsive load bids (as these are higher priority).

Buy bids for the unresponsive load have a high price attached, ensuring that they are always fulfilled if there is enough power. This approach is crucial as unresponsive loads cannot be switched off. There is no special algorithm for bid formulation - the unresponsive load is predicted by measurement, and a fixed high-price bid is created.

\subsubsection{EV Load and Supply}
EVs can act as either a load or a supply. They receive $\text{load}_\text{min}$ and $\text{load}_\text{max}$ publications from the EV federate (described later). A positive value of these variables indicates capacity for \textit{charging} load, whereas a negative value indicates capacity for \textit{discharging} power.
            At each market period, the EV first determines its load range $[\text{load}_\text{min}, \text{load}_\text{max}]$ according a simple algorithm which is designed to ensure that the state-of-charge (SoC) of the battery always stays within an acceptable range that will allow the owner to drive it. The algorithm determines the range of loads an EV will accept based on its SoC, location, and the time until it moves. If the EV is not at home or is leaving home before the next market period, it accepts no load changes. If the SoC is above 90\%, it can only discharge; between 30\% and 90\%, it can both charge and discharge; between 20\% and 30\%, it can only charge. Below 20\%, it is set to charge at its maximum rate.
    
            
            \begin{algorithm}
    \caption{EV Bid Formulation Procedure}
    \begin{algorithmic}
    \Procedure{FormulateEVBids}{}
        \State $\texttt{ma\_long} \gets$ Mean LMP over last 24 hours
        \State $\texttt{ma\_short} \gets$ Mean LMP over last 30 minutes
        \State $\texttt{iqr\_long} \gets$ IQR of LMP of last 24 hours
        \State $\texttt{ev\_buy\_price} \gets \texttt{ma\_long}$
        \State $\texttt{ev\_sell\_price} \gets \max(\texttt{ev\_buy\_price} + \texttt{iqr\_long} * 0.05, \texttt{ma\_short} + \texttt{iqr\_long} * 0.1)$
    \EndProcedure
    \end{algorithmic}
      \label{alg:ev_bid_formulation}
\end{algorithm}

            Next, the EV determines its market bids. If $\text{load}_\text{min}>0$ (i.e. the EV must charge), it submits only a buy bid for $\text{load}_\text{min}$ with a very high price. If $\text{load}_\text{max}<0$ (the EV must discharge), it submits only a sell bid for $\text{load}_\text{max}$ with a very low price. If the EV can either charge or discharge, it submits two bids: a buy order for $\text{load}_\text{max}$ and a sell order for $-\text{load}_\text{min}$.
            
            In the two-bid case, prices are determined according to the Algorithm \ref{alg:ev_bid_formulation},
            which is dependent on averages of the recent LMP history of the market. This pricing algorithm is simple but has a couple of important features:
            \begin{itemize}
                \item The sell price is always at least slightly above the buy price, which prevents the EV from pointlessly purchasing its own sell bids.
                \item The buy price tracks the 24-hr mean of the LMP, so the EV will buy when the price is relatively low and sell when it is relatively high.
            \end{itemize}

\subsubsection{PV Supply}
PV installations always aim to generate at their maximum capacity and sell all of their generated power, as any time spent generating below max capacity represents a missed opportunity for renewable power generation. PV power output is predicted by multiplying the voltage and current values for the PV array published from GridLAB-D, yielding wattage. The PV sell price is fixed at a value that creates a good market equilibrium where PV power wastage is minimized, and in this study we have used 0.0148\$/kWh.

\subsection{Co-simulation Development}
Co-simulation is a technique for simulation of multiple interconnected components in a synchronized manner. It facilitates the integration of different simulation models, often from different domains, to create a cohesive virtual environment for complex systems. By simulating these systems together, co-simulation allows for the examination and measurement of their mutual influence, feedback loops, and overall system performance. 
            The \emph{co-simulation architecture} in this work is based on PEMT-CoSim (Section \ref{sect_pemt}); however, it has been extensively modified by:
            \begin{itemize}
              \item Developing and integrating the novel EV federate.
              \item Developing and integrating the continuous double auction market (CDA) which replaces the incomplete non-continuous double auction in PEMT-CoSim.
              \item Redesigning the energy trading system such that all energy consumed must be traded on the market, and packets are defined by time rather than power.
              \item Comprehensively modernizing and rewriting the codebase and upgrading its dependencies to build a platform that is substantially more legible, efficient and extensible.
            \end{itemize}
                        
            

\subsubsection{EV Federate}
\begin{figure}[h]
    \centering
    \includegraphics[scale=0.2]{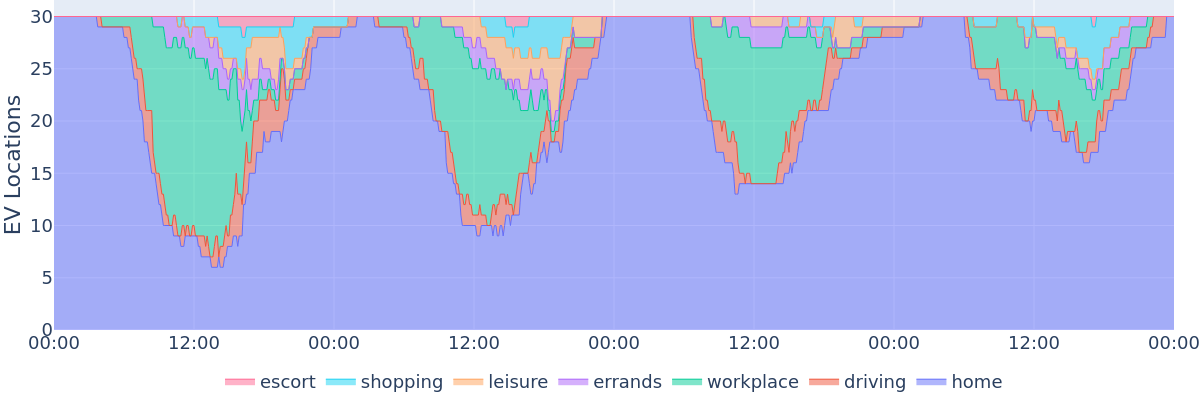}
    \caption{Locations of 30 EVs over a sample four-day period.}
    \label{EV_mob}
    \vspace{-1em}
\end{figure}

\begin{figure}[h]
    \centering
    \includegraphics[scale=0.21]{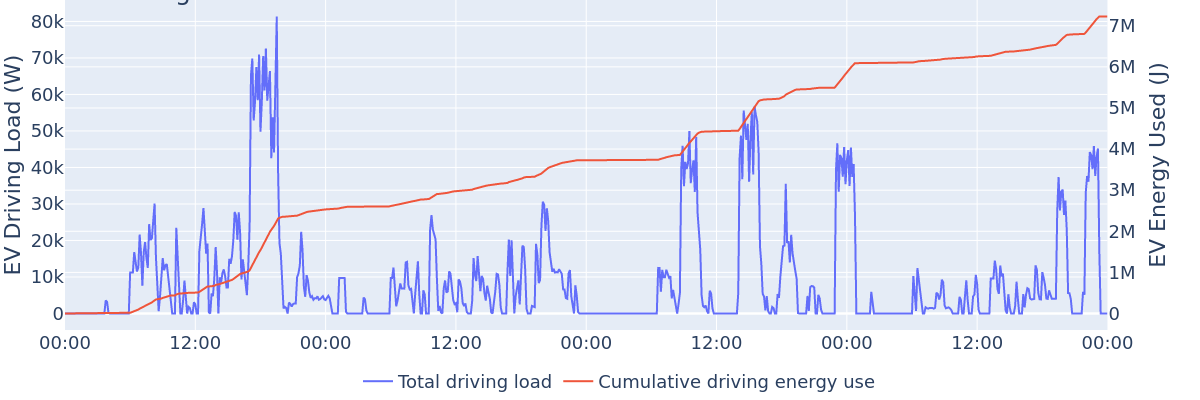}
    \caption{Battery discharge due to driving for 30 EVs (sample four-day period).}
    \label{EV_load}
\end{figure}

The function of the EV Federate is to incorporate EV mobility data and the resulting availability of
each EV to the microgrid for charging or discharging in the simulation. The mobility pattern of each
EV is generated using the emobpy Python library \cite{EV_lib} before the simulation runs. This library was selected as the data source as it is capable of generating realistic time-series mobility patterns based on recent documented vehicle mobility statistics and physical properties of battery-electric vehicles. The generated data includes the location, state, and distance driven of each EV over time and the corresponding battery load. A sample of four days of EV mobility data for 30 EVs is shown in Fig. \ref{EV_mob}.
In addition to mobility patterns, the emobpy library provides essential parameters related to
different car models, such as battery capacity, motor efficiency and maximum charging and discharging rates. These parameters are used to model the the battery discharge from driving over time. A sample of four days of total driving battery load data for 30 EVs and the resulting total cumulative energy
usage is provided in Fig. \ref{EV_load}.

Two per-vehicle variations in physical and mobility parameters contribute to a more diversified
and realistic simulation. 
\begin{itemize}
\item \textit{Car model} -- We incorporate two EV models available in emobpy: the Tesla Model Y
Long Range AWD and the Volkswagen ID.3. These models have been chosen based on their global
popularity, being among the highest-selling\footnote{https://cleantechnica.com/2023/03/03/best-selling-electric-cars-in-the-world-january-2023/} EVs in the world. 

\item \textit{Mobility rules} -- emobpy allows specification of rule sets which, alongside the statistical distributions,
govern the mobility profiles generated. We use a mixture of standard \emph{full-time worker}
and \emph{unemployed} rule sets in  80:20 ratio. 
\end{itemize}

The communication between the EV federate and other federates is structured around two sets of message types: subscriptions and publications. The EV federate subscribes to the desired charge rate from PET, and publishes its minimum and maximum load as well as its charging load to PET and GridLAB-D respectively. This interaction occurs in each market round.


%
%
%
%
%
%
%
%
%
%
%

\subsubsection{PET Market Mechanism}
Another key enhancement made to PEMT-CoSim is development of the PET market mechanism. The PET market in PEMT-CoSim utilized a double auction mechanism with
a single clearing price. This simple approach can limit the efficiency of the market and the number of orders matched in each round, especially when range of bid and ask prices is large.

To address these limitations, the market component of the substation federate was replaced with
an implementation of CDA. Unlike its predecessor, the CDA strives to
match as many orders as possible and permits transactions to take place at varying prices within the
same round. This feature is highly advantageous in a peer-to-peer PET trading scenario where the
availability of resources is time-bounded. For instance, if a solar PV installation fails to fulfill its sell
orders within a certain time frame due to the single clearing price constraint, the renewable power
generation opportunity is irreversibly lost. The CDA, by enabling more flexible
pricing, optimizes the matching of buy and sell orders, ensuring that resources such as renewable energy are utilized efficiently when available. The algorithm employed for this new market order matching is shown as Algorithm \ref{alg:order_matching}.

\begin{algorithm}[h]
\caption{Market Order Matching}
\begin{algorithmic}
\Procedure{MatchOrders}{$\texttt{bids}$}
    \State $\texttt{buyers} \gets$ Buyers from $\texttt{bids}$, sorted \(\downarrow\) by price 
    \State $\texttt{sellers} \gets$ sellers from $\texttt{bids}$, sorted \(\uparrow\) by price 
    \State $\texttt{transactions} \gets \emptyset$

    \While{$\texttt{buyers}$ \(\neq\) \(\emptyset\)}
        \State $\texttt{buyer} \gets$ $\texttt{buyers}[0]$
        \State $\texttt{matched} \gets$ first subset of sellers with net quantity  $\ge \texttt{buyer.quantity}$
        \If{$\texttt{matched} = \emptyset$}
            \State $\texttt{buyers.remove(buyer)}$
        \Else
            \For{$\texttt{seller}$ in $\texttt{matched}$}
                \State $\texttt{transaction\_{q}} \gets \min(\texttt{seller.quantity, buyer.quantity)}$
                \If{$\texttt{transaction\_q} > 0$}
                    \State Subtract $\texttt{transaction\_q}$ from buyer and seller quantities
                    \State Append transaction details to $\texttt{transactions}$
                    \If{$\texttt{seller.quantity} = 0$}
                        \State Remove $\texttt{seller}$ from $\texttt{sellers}$
                    \EndIf
                \EndIf
            \EndFor
            \If{$\texttt{buyer.quantity} = 0$}
                \State Remove $\texttt{buyer}$ from $\texttt{buyers}$
            \EndIf
        \EndIf
    \EndWhile

    \State \textbf{return} $\texttt{transactions}$
\EndProcedure
\end{algorithmic}
\label{alg:order_matching}
\end{algorithm}

\subsubsection{Codebase Improvements}
The PEMT-CoSim codebase was significantly improved through restructuring, modernization, and better readability. We hope to encourage other developers to use and contribute to the project. Key improvements made are outlined below. The updated codebase is available at https://github.com/0-ft/PEMT-CoSim.
            
            \textit{Code Refactoring} --
            The overall Python codebase was extensively refactored, tidied, and rationalized. This effort aimed at enhancing code quality, improving readability, structure, and maintainability. 

            \textit{Dependency Updating} --
            Several major dependencies were updated and necessary code changes made to reflect this. 

            \textit{Docker Containment} --
            The Docker container engine is used to host the simulation and its dependencies in a platform-independent environment. The Docker setup was rebuilt for cosimulation in the process of bringing the dependencies up-to-date, also upgrading the container operating system used. The new Docker setup  allows for single-run simulation containers to be created, allowing numerous concurrent simulations to be quickly spun up on a high-CPU host.

\subsection{Evaluation Metrics}
The key performance metrics for the PET scheme in each scenario are defined as follows. 

\subsubsection{Quality-of-Service (QoS)} 
This is the most important metric for the fitness of the energy trading system. We use house air temperature as a proxy for power sufficiency and indicator of QoS. Insufficient power results in deviations of air temperature from the setpoint, quantified by \(T_{\textrm{excess}}^2(t) = \max(T_{\textrm{air}} - T_{\textrm{setpoint}}, 0)^2\). A lower \(T_{\textrm{excess}}^2(t)\) suggests better power sufficiency. The average \(T_{\textrm{excess}}^2(t)\) across all houses at time \(t\) is \(\overline{T_{\textrm{excess}}^2(t)}\), with its time-averaged value \(\overline{T_{\textrm{excess}}^2}\) serving as the primary measure of power sufficiency.

\subsubsection{Supply Utilization}
This metric evaluates how well the PET system utilizes distributed energy resources such as PVs and EVs to reduce grid dependence. It measures the unused potential of these resources at any time \(t\) as \(P_{\textrm{surplus}}^{\textrm{resource}}(t)\), with the time-average \(\overline{P_{\textrm{surplus}}^{\textrm{resource}}}\) comparing against the average demand \(\overline{P_{\textrm{target}}}\).



%
%

\subsubsection{Energy Cost}
A core goal is to minimize the total cost for consumers, calculated by tallying the costs of accepted buy orders and the profits from sell orders. The cost effectiveness of the trading system is gauged using the volume-weighted average price (VWAP), which reflects the average transaction price, weighted by volume, over a specific time frame.

\section{Co-simulation Results}\label{sec_cosim_res}
We conduct co-simulations for different variations of the overall scenario in Fig. \ref{sim}. The number of houses, \(n_{\textrm{houses}}\), is 30 in all scenarios. The grid supply is uncapped in Scenario 1; in all other scenarios it is capped at 100 kW. The results are shown as time series behavior  over an average day. The values for each time are calculated as the average value at that time over four simulation days.
We define \(P_{\textrm{target}}(t)\) as the total power needed by the houses to meet their requirements at each time \(t\). The average of \(P_{\textrm{target}}(t)\) is given by \(\overline{P_{\textrm{target}}}\). The average power entering the system over time, i.e., \(\overline{P_{\textrm{supplied}}}\) must be at least \(\overline{P_{\textrm{target}}}\) or the supply is deficient. This metric becomes useful when complex time-varying supply dynamics make it difficult to see whether shifting of supply to different times (e.g. using EVs) could fix a shortfall.

\begin{itemize}
\item  \textbf{Scenario 1} -- \(n_{\textrm{EV}}=0\), \(n_{\textrm{PV}}=0\), \(\overline{T_{\textrm{excess}}^2}=0.2724\), \(\overline{\textrm{VWAP}}=0.01746\) USD/kWh, \(\overline{P_{\textrm{supplied}}}=\overline{P_{\textrm{target}}}=80\) kW
\end{itemize}

\begin{figure}
    \centering
    \subfigure[]{\includegraphics[width=0.44\textwidth]{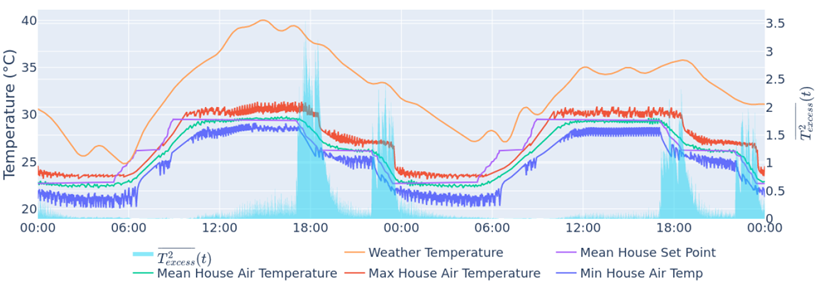}} 
    \subfigure[]{\includegraphics[width=0.44\textwidth]{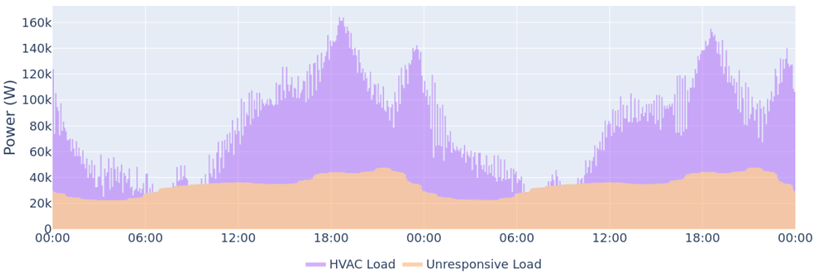}} 
    \subfigure[]{\includegraphics[width=0.44\textwidth]{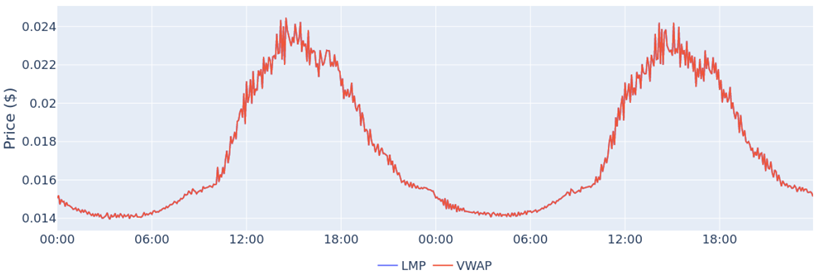}}
    \caption{Scenario 1: (a) HVAC performance; (b) load breakdown; (c) price.}
    \label{S1_perf}
    \vspace{-1.7em}
\end{figure}


Scenario 1 serves as a baseline, featuring only the main grid as the unlimited power source at the locational marginal price (LMP), without EVs or PV installations.
Fig. \ref{S1_perf} displays the time series of HVAC settings and temperatures: mean HVAC setpoint, house air temperatures (min, max, mean), and outdoor temperature alongside the metric \(\overline{T_{\textrm{excess}}^2(t)}\). With consistent power supply, HVAC maintains air temperature close to the setpoint, except for mornings when setpoint rises faster than the HVAC can heat, not affecting \(\overline{T_{\textrm{excess}}^2(t)}\) which remains low due to negligible positive deviations. Load fluctuations are significant; the highest load peaks at 18:30 and a lesser one at 23:30, correlating with HVAC setpoint reductions, not temperature highs. The lowest load at 06:00 coincides with minimal appliance use and the day’s coolest weather. The average power consumption of the microgrid, calculated using the trapezoidal rule, totals 80 kW from unresponsive loads (34.5 kW) and HVAC (45.5 kW). The average price mirrors the LMP, reflecting the grid’s role as the sole power supplier.

\begin{itemize}
\item \textbf{Scenario 2} -- \(n_{\textrm{EV}}=0\), \(n_{\textrm{PV}}=0\), \(\overline{T_{\textrm{excess}}^2}=3.335\), \(\overline{\textrm{VWAP}}=0.01726\) USD/kWh
\end{itemize}

\begin{figure}
    \centering
    \includegraphics[width=0.44\textwidth]{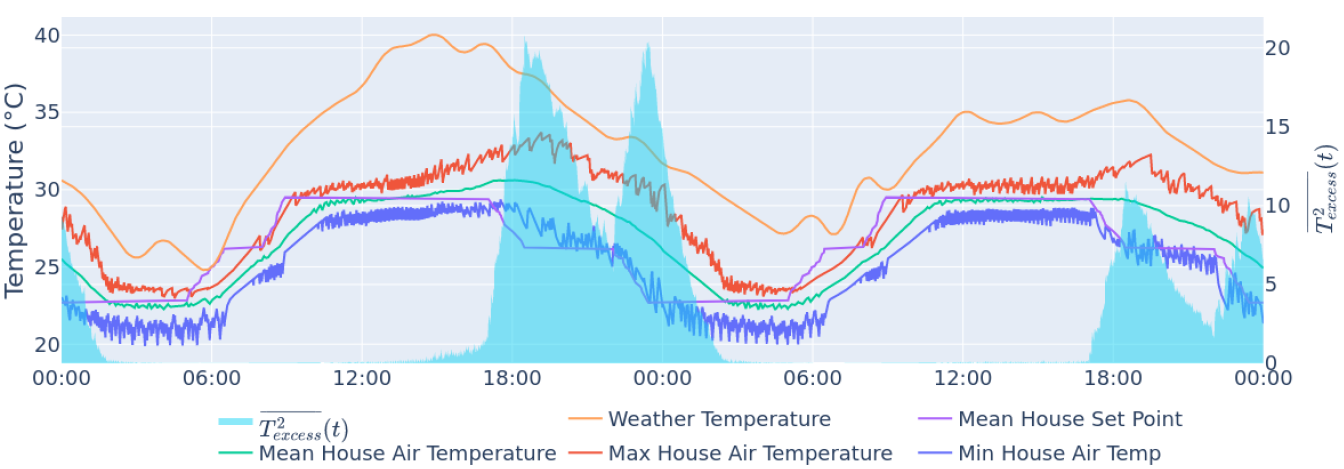}
    \caption{HVAC performance in Scenario 2. }
    \label{S2_perf}
    \vspace{-1.7em}
\end{figure}

Scenario 2 simply illustrates the impact of insufficient grid supply (capped at 100 kW) on the HVAC performance which is shown in Fig. \ref{S2_perf}. When the sum of the unresponsive load and the desired HVAC load exceeds the grid supply, house HVAC units shut down. This load shedding behaviour is most pronounced in the late afternoon and evening. As observed in Scenario 1, this period corresponds with the time when the HVAC load needed
to maintain the setpoint reaches its highest levels.
Consequently, the indoor air temperature deviates above the setpoint due to the lack of sufficient
HVAC operation. This deviation is clearly visible between approximately 17:00 and 02:00 illustrating the impact on QoS. For this reason, \(\overline{T_{\textrm{excess}}^2}\) is much higher than the value observed in Scenario 1.

\begin{figure}
    \centering
    \subfigure[]{\includegraphics[width=0.44\textwidth]{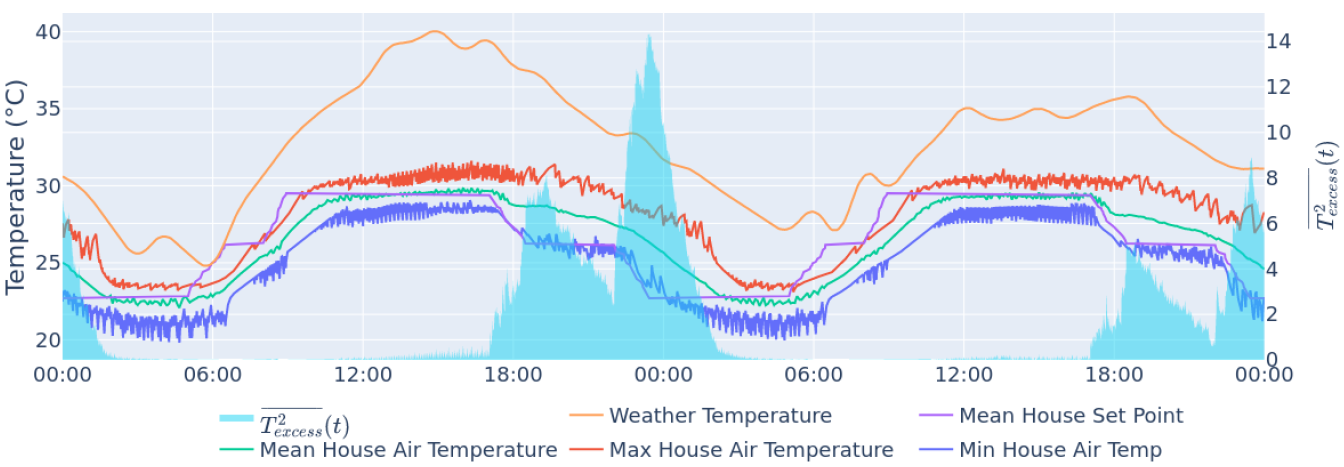}} 
    \subfigure[]{\includegraphics[width=0.44\textwidth]{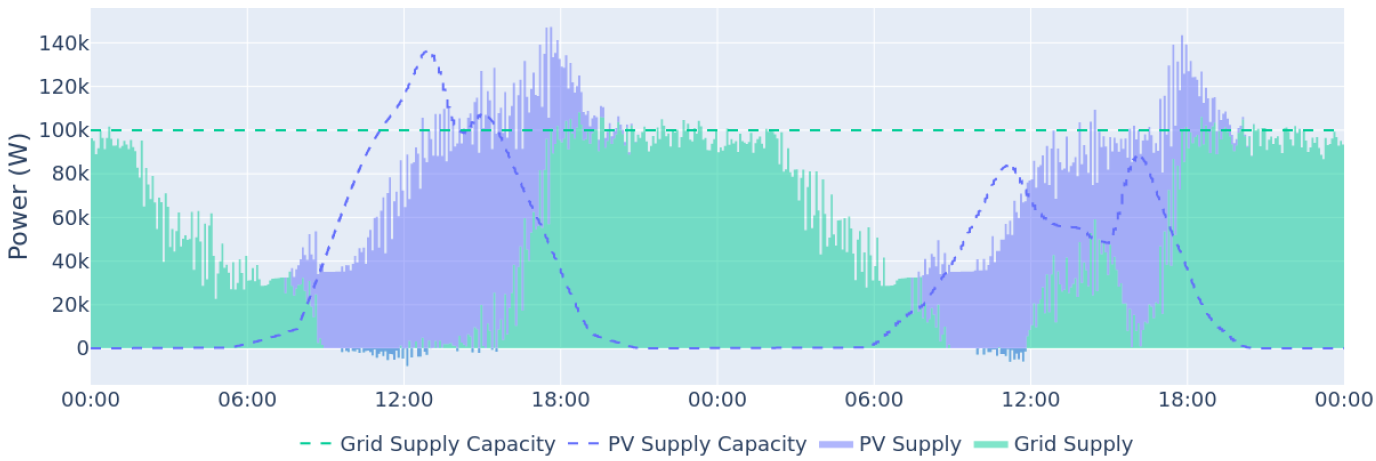}} 
    \subfigure[]{\includegraphics[width=0.44\textwidth]{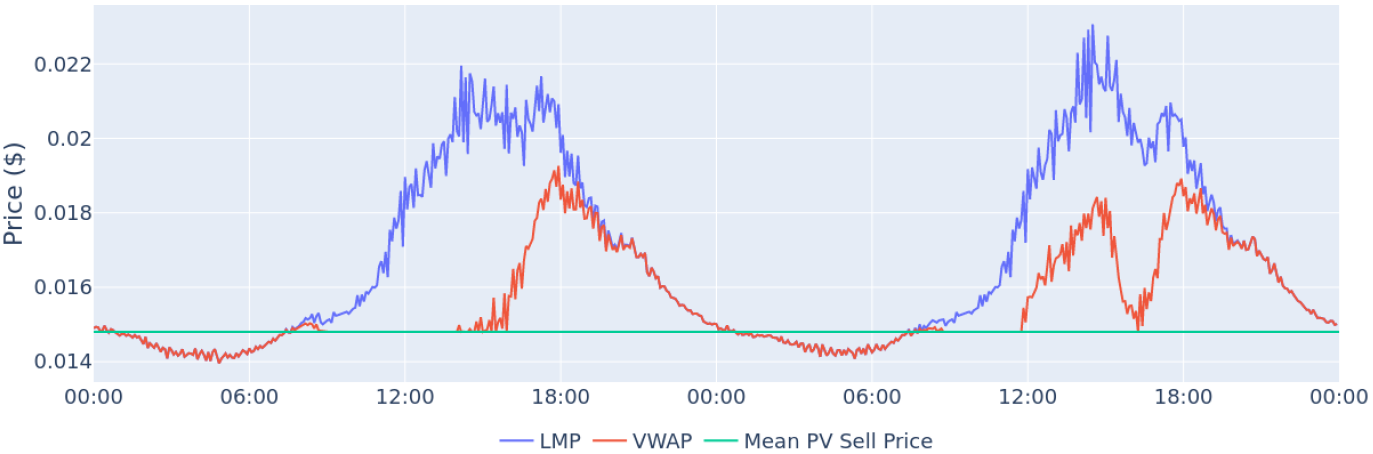}}
    \caption{Scenario 3: (a) HVAC performance; (b) supply breakdown; (c) price.}
    \label{S3_perf}
    \vspace{-1.7em}
\end{figure}

\begin{itemize}
\item \textbf{Scenario 3} -- \(n_{\textrm{EV}}=0\), \(n_{\textrm{PV}}=30\), \(\overline{T_{\textrm{excess}}^2}=1.737\), \(\overline{\textrm{VWAP}}=0.01556\) USD/kWh
\end{itemize}

In Scenario 3, all houses have solar PV installations, none have EVs, and the grid supply cap is 100 kW. This scenario depicts the case where the average power entering the system over a day exceeds \(\overline{P_{\textrm{target}}}\); however, due to timing mismatch of PV generation and power demand, there is still considerable excess temperature. Fig. \ref{S3_perf} shows the time series behavior of Scenario 3. The average power supply for the microgrid is the sum of the averages over time of the grid and PV capacity curves. Therefore, \(\overline{P_{\textrm{supplied}}}=\overline{P_{\textrm{supplied}}^{\textrm{PV}}}+\overline{P_{\textrm{supplied}}^{\textrm{grid}}}\)= 100 kW + 32.2 kW = 132.2 kW. 
The PV installations
can produce power from around 07:00 and reach their peak capacity around 12:00, but
their capacity diminishes through the afternoon and becomes negligible by 21:00. On the other
hand, the demand for power peaks at 18:30 and surpasses the grid cap from 13:00 right
through to 02:00. Therefore, despite the average power available being above the target, there
is underutilization of the PV resource, the average of which is given by  \(\overline{P_{\textrm{surplus}}^{\textrm{PV}}}\) = 5.275 kW. 
This underutilization results in excess temperature in the evening and night time, with a high
\(\overline{T_{\textrm{excess}}^2}\)
value of 1.73704 observed in the HVAC performance. This value is lower than that observed in Scenario 2, indicating  mitigating
effect from the PV installations (evident in the reduction of the first \(\overline{T_{\textrm{excess}}^2(t)}\) peak around 18:30,
and corresponding total power supply above the grid cap). However, it remains relatively high due to
the timing mismatch between PV generation and power demand.
The price plot reveals that during bright sunlight hours, the VWAP paid for power drops below the LMP. This is because the PV sell orders, which are priced below the LMP, fulfill some buy orders, effectively reducing the average price paid  during these hours. The overall average VWAP is therefore lower than that in Scenarios 1 and 2.

\begin{itemize}
\item \textbf{Scenario 4} -- \(n_{\textrm{EV}}=30\), \(n_{\textrm{PV}}=0\), \(\overline{T_{\textrm{excess}}^2}=0.2713\), \(\overline{\textrm{VWAP}}=0.01739\) USD/kWh
\end{itemize}

\begin{figure}
    \centering
    \subfigure[]{\includegraphics[width=0.44\textwidth]{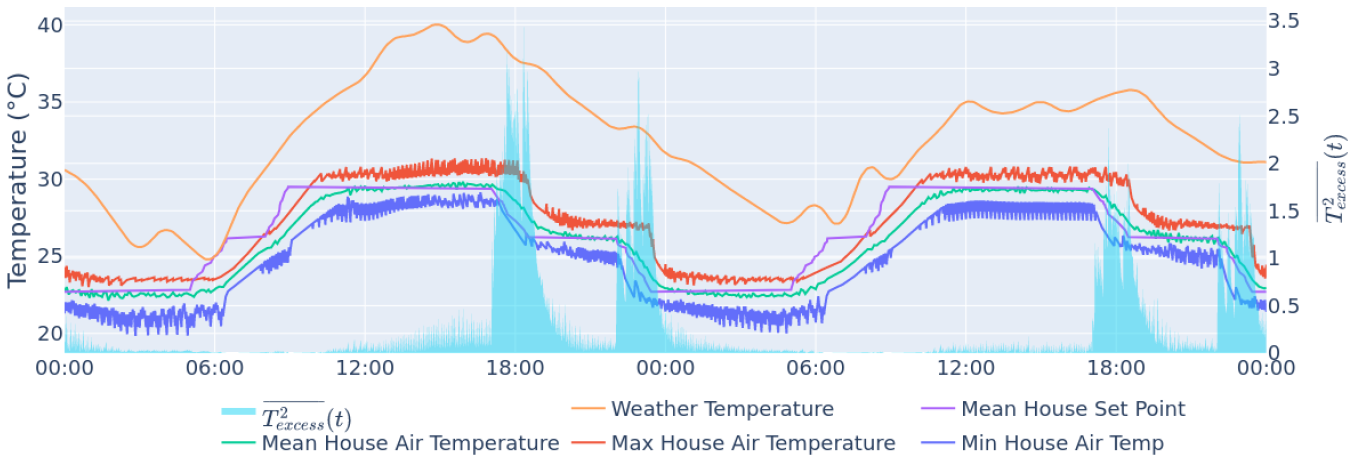}} 
        \subfigure[]{\includegraphics[width=0.44\textwidth]{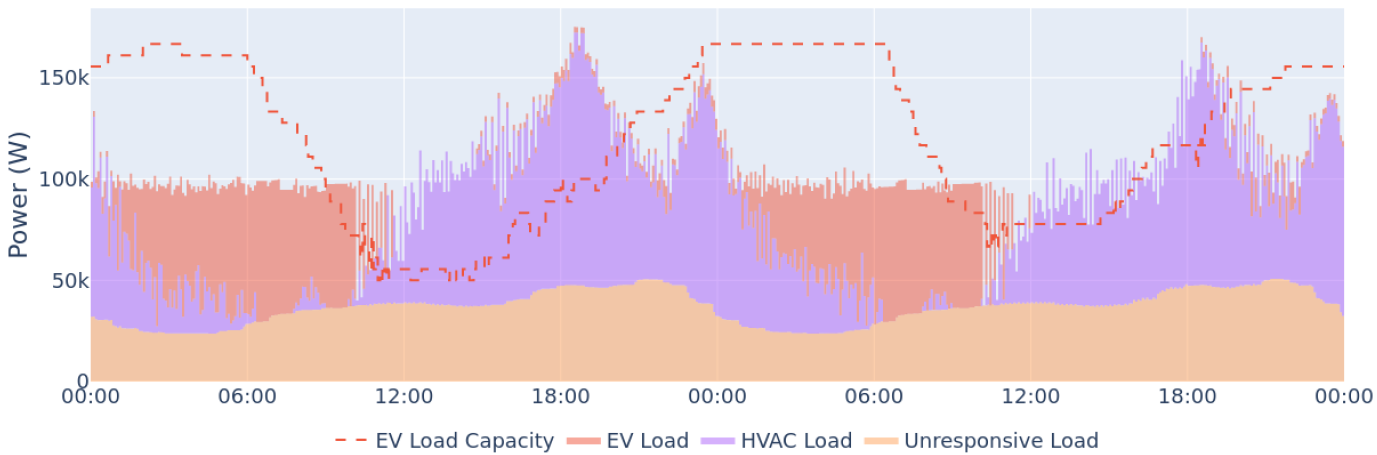}} 
    \subfigure[]{\includegraphics[width=0.44\textwidth]{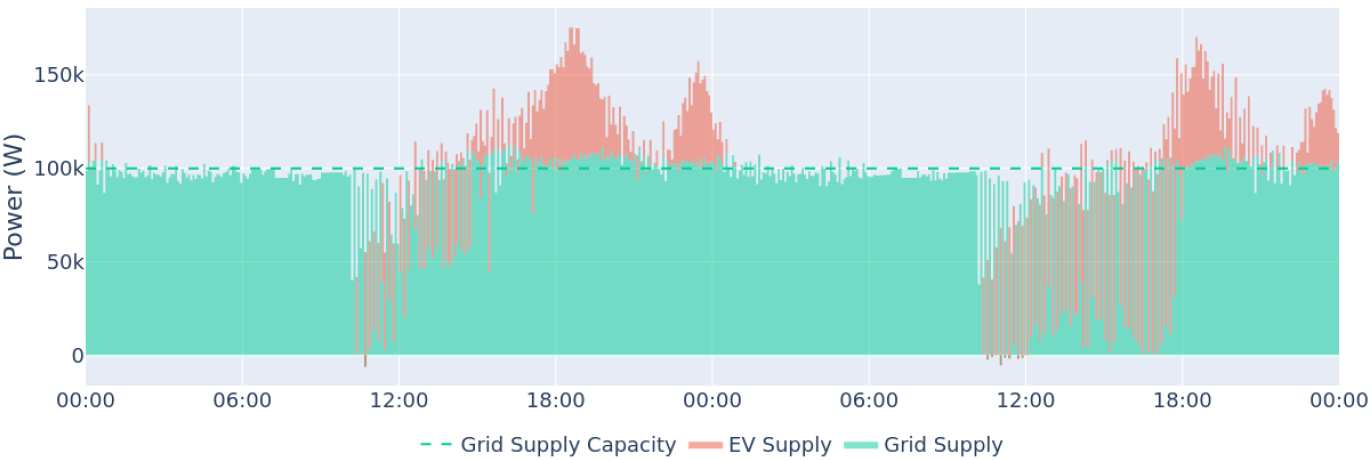}} 
    \subfigure[]{\includegraphics[width=0.44\textwidth]{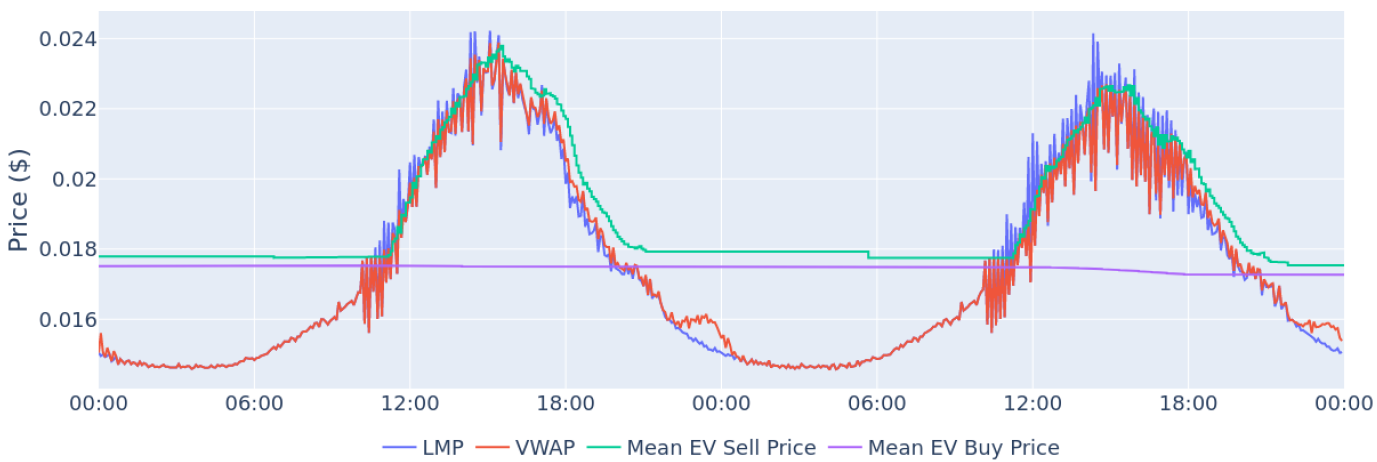}}
    \subfigure[]{\includegraphics[width=0.44\textwidth]{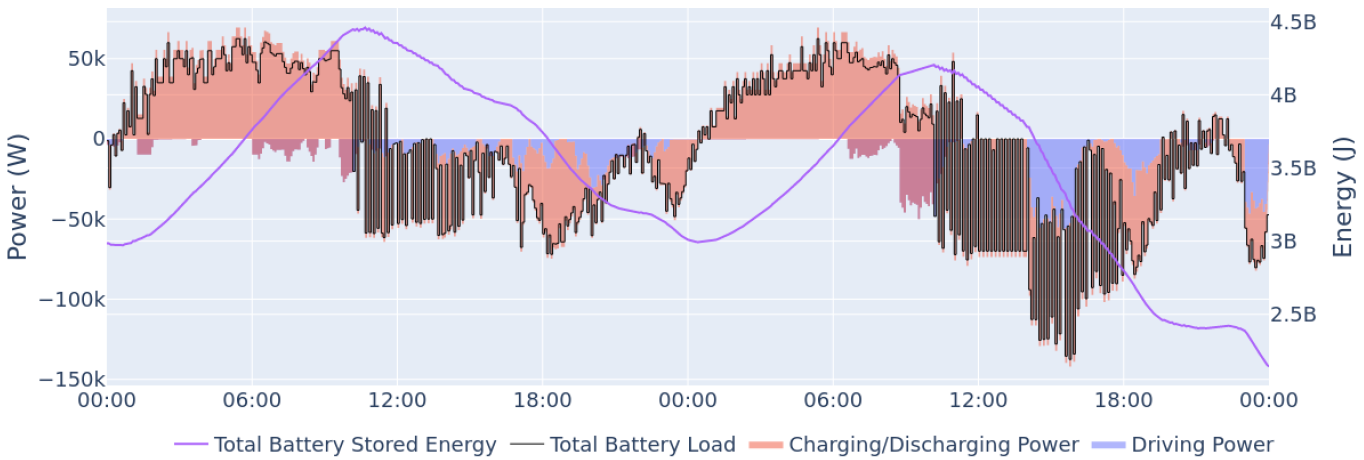}}
    \caption{Scenario 4: (a) HVAC performance; (b) load breakdown; (c) supply breakdown; (d) price; (e) EV battery and power state. }
    \label{S4_perf}
    \vspace{-1.7em}
\end{figure}

The microgrid in Scenario 4 incorporates 30 EVs and no PV installations, and maintains the grid
supply cap at 100 kW. This scenario demonstrates the role of EVs in flattening the load curve to match grid capacity and effectively utilizing the available power supply from the grid.
Fig. \ref{S4_perf} shows the performance in Scenario 4. The load and supply plots  clearly show the efficacy with
which the EVs match the supply curve to correspond to the required load. The EVs charge predominantly
between 01:00 and 10:00, when the LMP is low and most EVs are at home, and discharge
during times when the power demand surpasses the grid capacity. Discharge peaks around 19:00.
The load curve is substantially flattened, reflecting  effective load management aided by EVs.
The mean house air temperature closely adheres to the mean setpoint throughout
the entire day as shown in the HVAC behavior. The excess temperatures observed in Scenarios 2 and 3 during the early afternoon
and evening are significantly mitigated - peaks occur at similar times but with much lower values, similar to Scenario 1. \(\overline{T_{\textrm{excess}}^2}\) in this scenario is reduced to 0.2713, a significant improvement compared to
Scenarios 2 and 3. This is close to the ideal value in Scenario 1 and indicates that
the integration of EVs substantially improves the  ability of the system to maintain desired temperature.
The price plot  shows that the VWAP stays in close proximity to the LMP. This
behavior is anticipated as all power entering the system is procured at the LMP. The integration of
EVs does not  significantly alter the price paid, but their role in load-shifting enables better
fulfillment of power requirements at the same price point as Scenario 1.
Fig. \ref{S4_perf} also shows the total energy stored in EV batteries over time, as well as the change in this total energy. The most prominent peaks in driving load occur around 09:45 and 18:00.

\begin{itemize}
\item \textbf{Scenario 5} -- \(n_{\textrm{EV}}=30\), \(n_{\textrm{PV}}=30\), \(\overline{T_{\textrm{excess}}^2}=0.2616\), \(\overline{\textrm{VWAP}}=0.01565\) USD/kWh
\end{itemize}

\begin{figure}
    \centering
    \subfigure[]{\includegraphics[width=0.44\textwidth]{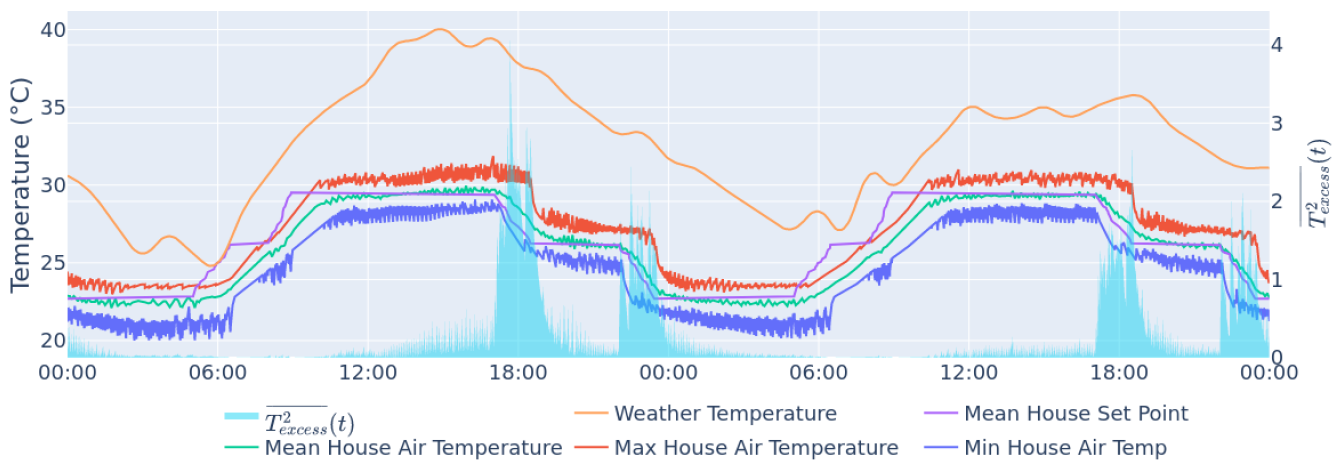}} 
        \subfigure[]{\includegraphics[width=0.44\textwidth]{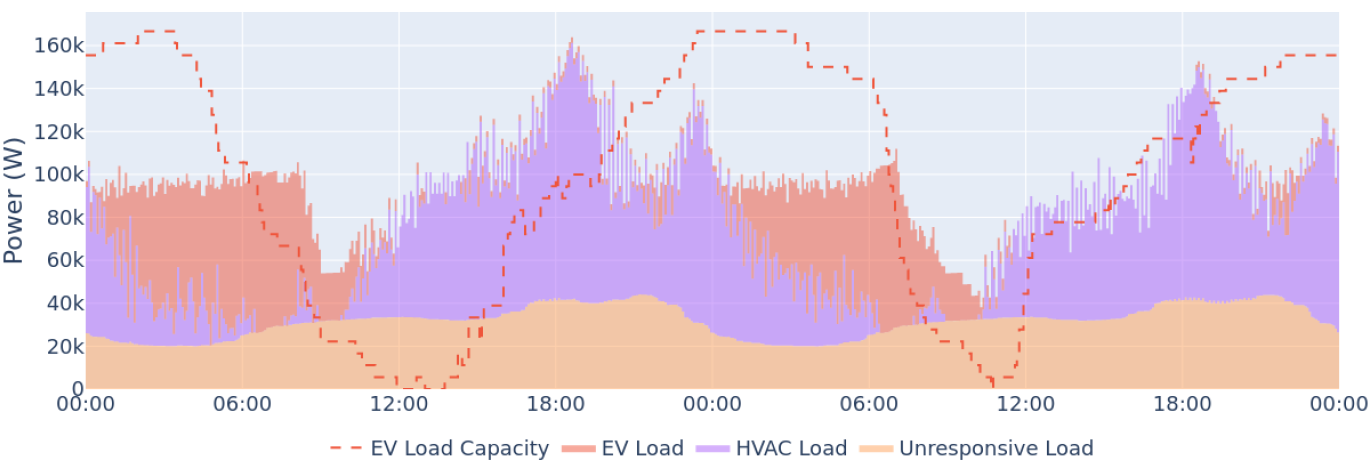}} 
    \subfigure[]{\includegraphics[width=0.44\textwidth]{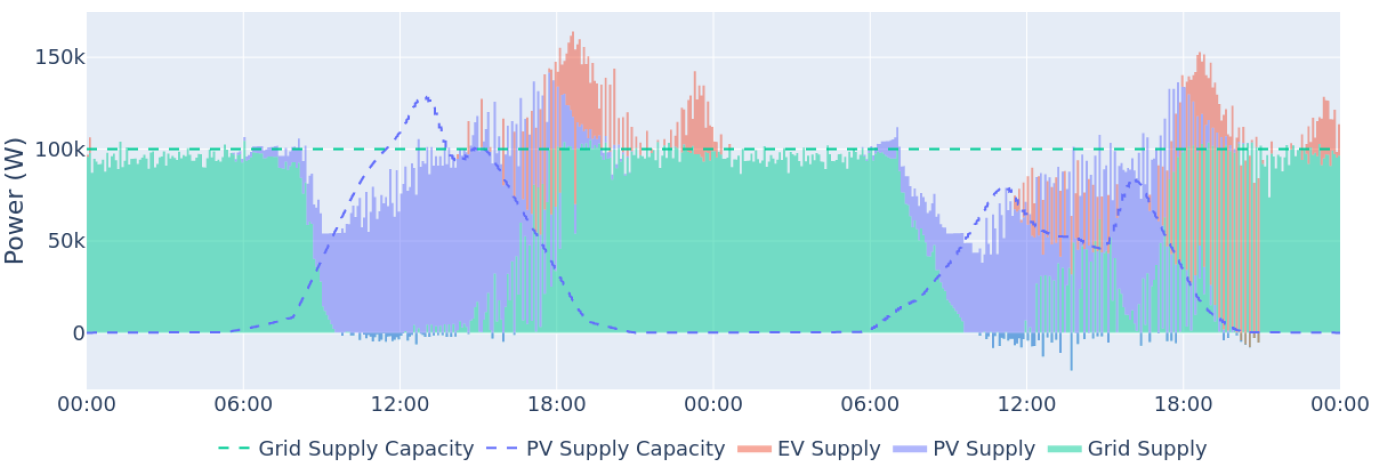}} 
    \subfigure[]{\includegraphics[width=0.44\textwidth]{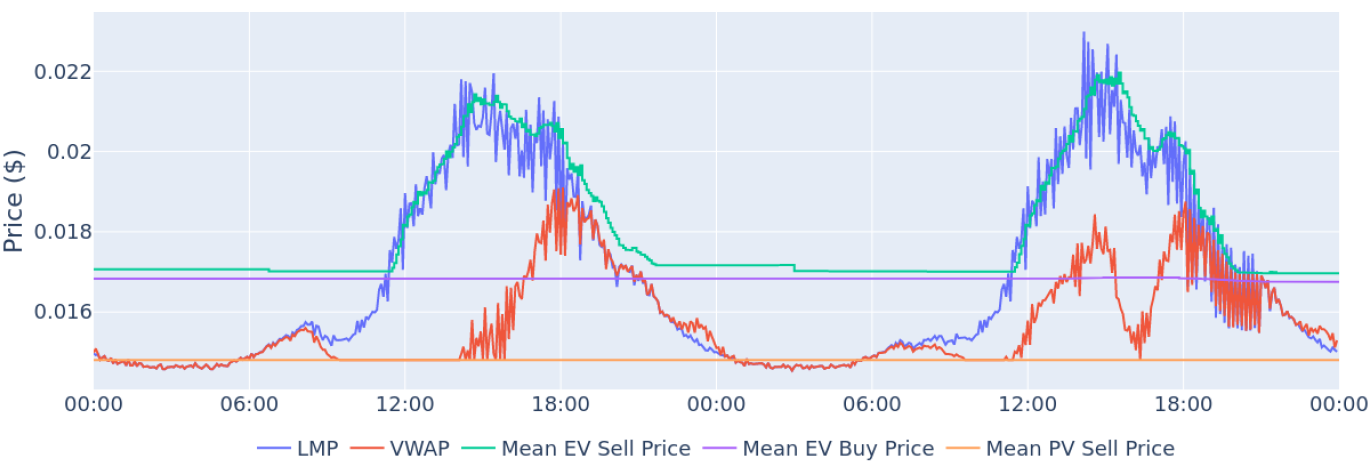}}
    \subfigure[]{\includegraphics[width=0.44\textwidth]{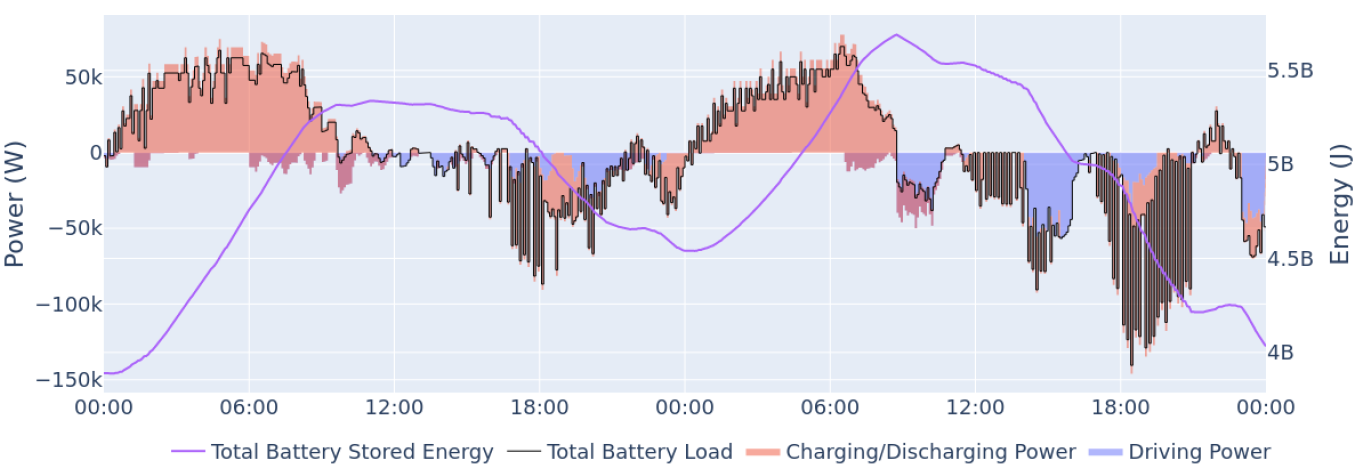}}
    \caption{Scenario 5: (a) HVAC performance; (b) load breakdown; (c) supply breakdown; (d) price; (e) EV battery and power state. }
    \label{S5_perf}
    \vspace{-1.7em}
\end{figure}

In Scenario 5, every house is equipped with a PV installation and an EV while the grid supply is capped at 100 kW. This scenario demonstrates
the potential of combining EVs with intermittent PV sources in the PET scheme,
demonstrating the collaborative behavior to meet  demand.
Fig. \ref{S5_perf} shows the performance in this scenario. The mean air temperature in the houses closely follows the mean setpoint
throughout as the HVAC system is effectively maintaining the desired temperature,
resulting in a very low value for \(\overline{T_{\textrm{excess}}^2}\), similar to Scenario 1. 
The load and supply curves in this scenario show that by acting as a reservoir for energy, the EVs ensure that the unused PV capacity  is
kept to a minimum. This is particularly evident during the morning hours from around 06:00 onwards,
when the supply of PV power is high, but HVAC demand is relatively low as the mean setpoint is
increasing. During this period, the EVs buy a substantial amount of power generated by the PV
installations. This results in the amount of power generated by the PV installations staying close to their maximum capacity, where without the EVs (as in Scenario 3) it would fall short. The EVs sell power mostly in the evening, starting around 18:00, when HVAC load is
high due to falling temperature set points, PV supply is low, and the grid supply is at capacity. By selling the stored power back to the microgrid at these times, EVs play a crucial role in meeting
demand. The average power price, \(\overline{\textrm{VWAP}}\), is also lower than that of all other scenarios, with the exception
of Scenario 3, which achieves a lower price but does not adequately meet the energy demands. The EV battery state is also shown in Fig. \ref{S5_perf}.

\section{Concluding Remarks}
The aim of this work was to develop and assess a co-simulation platform that could effectively demonstrate the utility of V2G EVs in microgrid PET schemes, particularly under power generation fluctuations. The co-simulation platform was successfully developed, and the outcomes from various simulations have indeed demonstrated that EVs, as part of a PET scheme, have the capacity to
significantly contribute to the smooth and continuous energy supply. The simulated PET scheme enables the utilization of a collection of EVs as a power reservoir.
While the availability of these vehicles to the microgrid is not constant,
the PET system manages this availability responsively. Coordinated EV charging/discharging
allows for substantial reduction of PV power underutilization while maintaining QoS for
and optimizing price paid by consumers. These findings are significant in the broader context of promoting the energy transition and tackling
climate change. Integration of V2G EVs and PET makes renewable energy sources more viable by ensuring a reliable and efficient power supply. 



\bibliographystyle{IEEEtran}
\bibliography{references.bib}

\end{document}